\documentclass[conference]{IEEEtran}
\normalsize

\usepackage{graphicx,amsmath,epsfig,times}
\usepackage{psfrag,cite}
\usepackage{amstext,amssymb,latexsym}
\usepackage{setspace}
\usepackage{algorithmic, algorithm}
\usepackage{booktabs,array}
\usepackage{caption}
\usepackage{amsmath}
\usepackage{multicol}

\usepackage[cmintegrals]{newtxmath}

\allowdisplaybreaks

\newcommand{\subparagraph}{}
\usepackage[compact]{titlesec}
\titlespacing{\section}{0pt}{1ex}{1ex}
\titlespacing{\subsection}{0pt}{1ex}{1ex}
\titlespacing{\subsubsection}{0pt}{1ex}{1ex}
\newcommand{\tableequation}[1]{%
  \vspace*{-\baselineskip}
  {\begin{flalign}#1&&&\end{flalign}}%
  \vspace*{-\baselineskip}
}

\begin{document}
%
% paper title
% can use linebreaks \\ within to get better formatting as desired
\title{Beamspace Aware Adaptive Channel Estimation for Single-Carrier Time-varying Massive MIMO Channels}

\author{\IEEEauthorblockN{Gokhan M. Guvensen and Ender Ayanoglu\\}
\IEEEauthorblockA{Center for Pervasive Communications and Computing (CPCC),
Dept. of EECS, UC Irvine, CA, USA\\
E-mail: guvensen@metu.edu.tr and ayanoglu@uci.edu}}

%\author{G\"{o}khan~M.~G\"{u}vensen,
%~\IEEEmembership{Member,~IEEE},~and~
%Ender~Ayano\u{g}lu,% <-this % stops a space
%~\IEEEmembership{Fellow,~IEEE}
%%\thanks{The authors are with the Department of Electrical Engineering and Computer Science, %University of California Irvine (UCI), CA, USA (e-mail: g.m.guvensen@uci.edu, ayanoglu@uci.edu.)}
%}

% use only for invited papers
%\specialpapernotice{(Invited Paper)}

% make the title area
\maketitle
\begin{abstract}
In this paper, the problem of sequential beam construction and adaptive channel estimation based on reduced rank (RR) Kalman filtering for frequency-selective massive multiple-input multiple-output (MIMO) systems employing single-carrier (SC) in time division duplex (TDD) mode are considered. In two-stage beamforming, a new algorithm for statistical pre-beamformer design is proposed for spatially correlated time-varying wideband MIMO channels under the assumption that the channel is a stationary Gauss-Markov random process. The proposed algorithm yields a nearly optimal pre-beamformer whose beam pattern is designed sequentially with low complexity by taking the user-grouping into account, and exploiting the properties of Kalman filtering and associated prediction error covariance matrices. The resulting design, based on the second order statistical properties of the channel, generates beamspace on which the RR Kalman estimator can be realized as accurately as possible. It is observed that the adaptive channel estimation technique together with the proposed sequential beamspace construction shows remarkable robustness to the pilot interference. This comes with significant reduction in both pilot overhead and dimension of the pre-beamformer lowering both hardware complexity and power consumption.
\end{abstract}

%\begin{keywords}

%IEEEtran, journal, \LaTeX, paper, template.
%\end{keywords}
% Note that keywords are not normally used for peerreview papers.

% For peer review papers, you can put extra information on the cover
% page as needed:
% \begin{center} \bfseries EDICS Category: 3-BBND \end{center}
%
% For peerreview papers, inserts a page break and creates the second title.
% Will be ignored for other modes.
%\IEEEpeerreviewmaketitle

\section{Introduction}
\label{sec:introduction}
%Massive multiple-input multiple-output (MIMO) systems, which are equipped with a large number of antenna elements at the base station (BS) to serve a relatively smaller number of user terminals (UTs) simultaneously, are believed to be one of the key technologies for next generation cellular systems such as the upcoming 5G standard \cite{Larsson14,andrews14}.
%With its potential large gains in spectral and energy efficiency, massive MIMO is especially promising for outdoor cellular systems operating at millimeter (mm) wave frequencies, where large antenna arrays can be packed into small form factors, and extremely large bandwidths are available for commercial use (e.g., up to 7 GHz in the 60 GHz band) \cite{ghosh14,ayanoglu14}. Thus, 
It is anticipated that massive MIMO systems in the mm wave range form an important part of 5G systems expected to support much larger (e.g., 1000 times faster) data rates than the currently deployed standards \cite{andrews14}.
%Instantaneous CSI at the base station (BS) is essential for massive MIMO transmission in order to capitalize the spatial diversity and multiplexing benefits of the channel \cite{swindlehurst14}.   
In practice, channel state information (CSI) is typically obtained with the assistance of the periodically inserted pilot signals \cite{swindlehurst14}. This brings the pilot overhead 
%, namely, the amount of transmission resources (signaling dimensions per time-frequency channel coherence slot) 
consumed by the training data to be proportional to the number of active users in the system for uplink training, and the number of BS antennas for downlink training respectively \cite{ashikhmin11}. 
%The acquisition of CSI in massive MIMO transmission has been studied extensively in literature. One of the primary framework is frequency division duplex (FDD) mode, where CSI is typically obtained through explicit downlink training and uplink (limited feedback) \cite{zoltowski14}. Since use of the FDD operation imposes a severe limit on the number of BS antennas due to the pilot overhead, alternatively, 
We assume 
CSI at the BS can be acquired by means of uplink training in time division duplex (TDD) mode, where the uplink pilots provide the BS with downlink as well as uplink channel estimates simultaneously via leveraging channel reciprocity \cite{swindlehurst14,ashikhmin11}.

The processing of the signals with very large dimensionality, the pilot interference, and the pilot overhead are thought to be limiting factors for an accurate channel acquisition and throughput of massive MIMO transmission in mm wave especially in high mobility or in applications requiring low latency and short-packet duration. 
Recently, the \textit{two-stage beamforming} concept under the name of Joint Spatial Division and Multiplexing (JSDM) \cite{adhikary13,nam14} has been proposed to reduce the dimension of the MIMO channel effectively, and to enable massive MIMO gains and simplified system operations \cite{kim15}. Even though JSDM is suggested as an effective reduced-complexity two-stage downlink precoding scheme for flat-fading multi-user MIMO systems in frequency division duplex (FDD) mode initially, the idea of \textit{two-stage beamforming} (in \cite{adhikary13,nam14,kim15}) can be applied to both downlink and uplink transmission in TDD. 
%JSDM can be seen as a divide-and-conquer approach considering the fact that the channel between a user and BS is spatially correlated. 
The key idea lies in \textit{user-grouping}, i.e., partitioning the user population supported by the serving BS into multiple groups each with approximately the same channel covariance eigenspaces. Then, one can decompose the MIMO beamformer at the BS into two steps via the use of a spatial \textit{pre-beamformer}, which distinguishes \textit{intra-group} signals from other groups by suppressing the \textit{inter-group interference} while reducing the signaling dimension. The major complexity reduction comes from the approach that the \textit{pre-beamformer} is properly designed based only on the second-order statistics of the channel, and not on the instantaneous CSI (which varies on a much higher rate). 

%In this case, the subsequent operations such as downlink multi-user precoding and uplink detection/decoding algorithms can be fulfilled based on the CSI of the \textit{effective} channel with significantly reduced dimensions thanks to the \textit{pre-beamforming} projection. At the same time, the training dimension necessary to learn the \textit{effective} channels of each user terminal (UT) is reduced considerably. 

%Also, the JSDM scheme motivates the use of analog/digital MIMO architectures, specifically the so called \textit{hybrid beamforming} \cite{ayach14,liu14,heath14,Noh16}, recently proposed as an alternative for fully digital precoding/decoding in mm wave, where efficient reconfigurable radio frequency (RF) architectures will be implemented at competitive cost, size, and energy in the near future. In the \textit{hybrid beamforming} architecture, the statistical \textit{pre-beamformer} (which depends on slowly varying parameters) may be implemented in the analog RF domain, while the multi-user MIMO precoding/decoding stage can be implemented by standard baseband processing.       

The wideband massive MIMO channel is expected to be sparse both in the angle and time (delay) domains.
The \textit{channel sparsity} \cite{adhikary14,swindlehurst16}, which becomes particularly relevant at mm wave frequencies, is observed in practical cellular systems, where the channels are often characterized with limited scattering and hence correlated in the spatial domain; the BS sees the incoming multi-path components (MPCs) under a very constrained angular range, i.e., angle of arrival (AoA) support, and the MPCs occur in clusters in the \textit{angle-delay} plane \cite{adhikary14}. Thus, it is important to design statistical \textit{pre-beamformer} yielding an efficient implementation of channel estimator taking this sparsity into account. Moreover, the subsequent stages preceded by the channel estimation can be realized in much lower complexity. 

In this paper, reduced rank Kalman filter based CSI estimator together with the sequential design of statistical \textit{pre-beamformer} is proposed for frequency-selective massive MIMO systems employing single-carrier (SC) modulation in TDD mode. With proposed pre-beamformer design, taking the Kalman prediction errors at each eigen-direction and \textit{inter-group} interference in JSDM into account, significant reduction in beamspace dimension is obtained, which reduces the pilot overhead and hardware complexity considerably for hybrid beamforming.   

\section{System Model} 
\label{sec:sym_model}
We consider a cellular system based on massive MIMO transmission operating at mm wave bands in the TDD mode employing SC in which a BS, having $N$ antennas, serves $K$ single-antenna UTs. In order to reduce the overhead while acquiring the instantaneous CSI associated with massive MIMO, \textit{two-stage beamforming} \cite{adhikary13,nam14,kim15} is adopted throughout this paper.  
%The main idea is based on partitioning the user population supported by the serving BS into multiple groups in order to enable massive MIMO gains and simplified system operations \cite{adhikary13,nam14}.
It is assumed that  
$K$ users are partitioned into $G$ groups, where the $K_g$ users in group $g$ have statistically independent but identically distributed (\textit{i.i.d.}) channels \cite{adhikary13,nam14}.

The frequency-selective massive MIMO channel is supposed to be time-varying in general. At the beginning of every coherence interval, all users of the intended group $g$ 
transmit training sequences with length $T$. We assume a linear modulation (e.g., PSK or QAM) and a transmission over frequency-selective channel for all UTs with a slow evolution in time relative to the signaling interval (symbol duration). Under such conditions, the baseband equivalent received signal samples, taken at symbol rate ($W$) after pulse matched filtering, are expressed as%
\footnote{
Only the UTs, belonging to same group, are assumed to be synchronized for coherent uplink SC transmission.}
\begin{align}
\mathbf{y}_n&=\underbrace{\sum_{\left\{k=1,\;g_k \in \Omega_g\right\}}^{K_g}\sum_{l=0}^{L_g-1}\mathbf{h}_{n,l}^{(g_k)}
x_{n-l}^{(g_k)}}_
{\textrm{Intra-Group Signal}} \nonumber \\
&\quad + \underbrace{\sum_{\left\{g'=1 
\; \vert g' \neq g\right\}}^G
\left(\sum_{\left\{k=1,\;g'_k \in \Omega_{g'}\right\}}^{K_{g'}}\sum_{l=0}^{L_{g'}-1}
\mathbf{h}_{n,l}^{(g'_k)}x_{n-l}^{(g'_k)}\right)+\mathbf{n}_n}
_{\boldsymbol{\eta}_n^{(g)}: \textrm{Inter-Group Interference + AWGN}}
\label{eqn:multiuser_multipath_mimo_model}
\end{align} 
\noindent for $n=0,1,2,\ldots$, where $\mathbf{h}_{n,l}^{(g_k)}$ is $N \times 1$ time-varying multi-path channel vector, namely, the array impulse response of the serving BS stemming from the $l^{th}$ multi-path component (MPC) of $k^{th}$ user in group $g$ at the  $n^{th}$ signaling interval. 
%It can be regarded as the discrete-time equivalent form of the channel response, and obtained after the symbol rate sampling of the impulse response, arising as the sum of the contributions from discrete MPCs, without loss of information. 
Here, 
$\left\{x_n^{(g_k)};\; n=-L_g+1,\ldots\right\}$ are the training symbols for the $k^{th}$ user in group $g$%
\footnote{
Training sequences are assumed to be non-orthogonal for synchronized \textit{intra-group} users for SC transmission in general.},
$L_g$ is the channel memory of group $g$ multi-path channels,
$\Omega_g$ is the set of all UTs belonging to group $g$ with cardinality $|\Omega_g|=K_g$, and $\left\{g_k\right\}_{k=1}^{K_g}$ are UT indices forming $\Omega_g$. The $L_g-1$ symbols at the start of the preamble, prior to the first observation at $n=0$, are the precursors. Training symbols are selected from a signal constellation $S \in \mathbb{C}$ and $\mathbb{E}\left\{|x_n^{(g_k)}|^2\right\}$ is set to $E_s$ for all $g_k$. 
In (\ref{eqn:multiuser_multipath_mimo_model}), $\mathbf{n}_n$ are the additive white Gaussian noise (AWGN) vectors during uplink pilot segment with spatially and temporarily 
\textit{i.i.d.} as $\mathcal{CN}\left(\mathbf{0},N_0\mathbf{I}_N\right)$, and $N_0$ is the noise power.
The first term of (\ref{eqn:multiuser_multipath_mimo_model})
is the received signal of the intended group $g$, named as the \textit{intra-group} signal of group $g$ users. The 
second term, $\boldsymbol{\eta}_n^{(g)}$, namely the \textit{inter-group interference}, comprises of all the interfering signals, which stem from all inner or outer cell users belonging to different groups other than $g$.
In (\ref{eqn:multiuser_multipath_mimo_model}), we assume users come in groups, either by nature or by the application of proper \textit{user grouping} algorithms in \cite{nam14,adhikary14}, which are out of scope of this work. 
Finally, the average received signal-to-noise ratio (\textit{snr}) can be defined as 
$snr \triangleq \frac{E_s}{N_0}$% 
\footnote{
It shows the maximum achievable \textit{snr} after beamforming when the beam is steered towards a point, i.e., angular location by assuming that the channel is normalized so that $\frac{1}{N}\frac{E_s}{N_0}$ can be seen as the average received \textit{snr} at each antenna element before beamforming.}.                     

\subsection{Fundamental Assumptions on Signal and Channel Model}
\label{sec:assump_model}
Each resolvable MPC of the users, belonging to any group $g$, is assumed to span some particular angular sector in azimuth-elevation plane, capturing local scattering around the corresponding UTs' angle of arrival (AoA) (with respect to the BS). Then, their corresponding cross-covariance  matrices can be expressed in the form of
\begin{equation}
\mathbb{E}\left\{\mathbf{h}_{n,l}^{(g_k)}\left(\mathbf{h}_{n,l'}^{(g'_{k'})}\right)^H\right\}=\rho_l^{(g)}\mathbf{R}_l^{(g)}\delta_{gg'}\delta_{kk'}\delta_{ll'}, 
%\nonumber \\
%\textrm{ where } & \sum_{l=0}^{L_g-1}\rho_l^{(g)}=1,\;
%\operatorname{Tr}\left\{\mathbf{R}_l^{(g)}\right\}=1 
\label{eqn:mimo_multipath_correlations}
\end{equation}
\noindent for 
$\sum_{l=0}^{L_g-1}\rho_l^{(g)}=1$ and 
$\operatorname{Tr}\left\{\mathbf{R}_l^{(g)}\right\}=1$ by using the uncorrelated local scattering model where all MPCs are assumed to be mutually independent according to the well-known wide sense stationary uncorrelated scattering (WSSUS) model \cite{adhikary14,swindlehurst16}, the multi-path channel vectors are uncorrelated with respect to $l$, and also mutually uncorrelated with that of the different users (independent of whether in the same group or not). 
In (\ref{eqn:mimo_multipath_correlations}), $\rho_l^{(g)}$ is the power delay profile (pdp) of the group $g$ multi-path channels, showing the average channel strength at each delay, and $\mathbf{R}_l^{(g)}$, with $r_{g,l}$ non-zero or dominant eigenvalues, can be considered as the common spatial covariance matrix of group $g$ UTs at $l^{th}$ delay. In this model, each antenna element at a BS is assumed to see the incoming MPCs at the same common support on the \textit{angle-delay} plane (similar to the one in 
\cite{you15,adhikary14}). Also, the effective rank of $\mathbf{R}_l^{(g)}$, namely, $r_{g,l}$ is expected to be much smaller than the number of array elements, $N$, due to the 
\textit{channel sparsity} pronounced at mm wave \cite{rappaport13,adhikary14}. 

The channel \textit{long-term} channel spatial correlation characteristic is assumed to be stationary. In fact, $\mathbf{R}_l^{(g)}$s are slowly varying in time as the AoA of each user signal evolves depending on the user mobility, variation rate of the scattering environment characteristics, etc.\cite{you15,adhikary14,utschick05}. However, this rate of change is significantly lower than that of the small-scale fading (instantaneous CSI), and they can be estimated with guaranteed accuracy for all intended groups in practice with the help of fast initial acquisition techniques or by acquiring the AoA sector of each user group \cite{you15}. Here, our focus is on the tracking of instantaneous CSI with significantly reduced complexity and overhead.   

When Rayleigh-correlated channel coefficients are assumed such
as $\mathbf{h}_l^{(g_k)}\thicksim 
\mathcal{CN}\left(\mathbf{0},\rho_l^{(g)}\mathbf{R}_l^{(g)}\right)$, mutually independent across the users for all $g_k$,   

Spatio-temporal covariance matrix of the \textit{inter-group interference} in (\ref{eqn:multiuser_multipath_mimo_model}) can be calculated by taking \textit{long-term} expectation over all MPCs $\mathbf{h}_{n,l'}^{(g'_{k'})}$s other than the ones belonging to group $(g)$ in spatial domain, and transmitted symbols $x_{n'}^{(g'_{k'})}$s in temporal domain. Considering the mutual independence across multi-path channel vectors (due to the WSSUS model) given by (\ref{eqn:mimo_multipath_correlations}), and considering that the transmitted symbols of different users are uncorrelated (including the data transmission period), i.e.,
$\mathbb{E}\left\{x_{n}^{(g_k)}\left(x_{n'}^{(g'_{k'})}\right)^H\right\}
=\gamma^{(g)}E_s\delta_{nn'}\delta_{gg'}\delta_{kk'}$, the following is obtained: $\mathbb{E}\left\{\boldsymbol{\eta}_n^{(g)}
\left(\boldsymbol{\eta}_{n'}^{(g)}\right)^H\right\}=
\mathbf{R}^{(g)}_{\boldsymbol{\eta}}\delta_{nn'}$, where 
\begin{equation}
\mathbf{R}^{(g)}_{\boldsymbol{\eta}} \triangleq 
E_s \left(\sum_{g' \neq g} \gamma^{(g')} K_{g'} \sum_{l=0}^{L_{g'}-1}
\rho_l^{(g')}\mathbf{R}_l^{(g')}\right)+N_0\mathbf{I}_N,
\label{eqn:InterGroupInterference} 
\end{equation}
\noindent and $\gamma^{(g')}$ for $g' \neq g$ can be regarded as the relative average received power at BS of \textit{inter-group} users normalized with that of the group $g$ users. In (\ref{eqn:InterGroupInterference}), 
$\gamma^{(g')}$s are accountable for the \textit{near-far} effect stemming from the fact that average received signal strength of different UTs may differ significantly depending on their distance to the BS. Moreover,      
It is important to note that $N \times N$ covariance matrix of the \textit{inter-group interference} $\mathbf{R}^{(g)}_{\boldsymbol{\eta}}$ in (\ref{eqn:InterGroupInterference}) consists of all the statistical information of the CSI in spatial domain (i.e., AoA support) for all inner and outer cell users interfering with group $g$ users.
%The interference covariance matrix, $\mathbf{R}^{(g)}_{\boldsymbol{\eta}}$ can be obtained by using the common spatial covariance of each intended group, namely, $\mathbf{R}_l^{(g)}$ in (\ref{eqn:mimo_MPC_covariance}) at each delay. 

\subsection{Spatio-Temporal Domain Vector Definitions}
\label{sec:space_time_def}
Before elaborating on the details of the estimation technique, we give the following matrix and vector definitions that will be useful in the subsequent sections. First, the training vector (or convolution vector), comprising of the transmitted pilots for $k^{th}$ user in group $g$ at the $n^{th}$ signaling interval, is defined as $\mathbf{x}_{n,k}^{(g)}\triangleq\left[x_n^{(g_k)} \cdots x_{n-L_g+1}^{(g_k)} \right]^H$. In a similar manner, the extended multi-path channel vector of the $k^{th}$ user, belonging to the intended group $g$, is given as
$\mathbf{f}_{n,k}^{(g)}\triangleq\left[\mathbf{h}_{n,0}^{(g_k)}
\cdots \mathbf{h}_{n,L_g-1}^{(g_k)}\right]^T_{NL_g \times 1}$ 
by concatenating all MPCs of the $k^{th}$ user in group $g$. Then, by using these defined ones, it will be useful to construct the following vector that represents the concatenated channel vector (including all the related channel parameters of users in group $g$ to be estimated simultaneously):   
\begin{equation}
\mathbf{h}_n^{(g)} \triangleq \operatorname{vec}\left\{\left[\begin{array}{cccc}
\mathbf{f}_{n,1}^{(g)} & \mathbf{f}_{n,2}^{(g)} & \cdots & \mathbf{f}_{n,K_g}^{(g)} \end{array}\right]_{NL_g \times K_g}\right\}. 
\label{eqn:vector_def2} 
\end{equation}
Finally, the complete training vector that consists of the training data of all users in group $g$ at the $n^{th}$ signaling interval is given by 
\begin{equation}
\mathbf{x}_n^{(g)} \triangleq \operatorname{vec}\left\{\left[\begin{array}{cccc}
\mathbf{x}_{n,1}^{(g)} & \mathbf{x}_{n,2}^{(g)} & \cdots & \mathbf{x}_{n,K_g}^{(g)} \end{array}\right]_{L_g \times K_g}
\right\}. 
\label{eqn:complete_training_matrix}
\end{equation}

Based on (\ref{eqn:vector_def2}) and (\ref{eqn:complete_training_matrix}), the spatio-temporal domain signal model is obtained by expressing $\mathbf{y}_n$ in (\ref{eqn:multiuser_multipath_mimo_model}) as
\begin{equation}
\mathbf{y}_n = \left(\mathbf{x}_n^{(g)} \otimes 
\mathbf{I}_N\right)^H\mathbf{h}_n^{(g)}+
\boldsymbol{\eta}_n^{(g)}
\label{eqn:space_time_equivalent_model}
\end{equation}

\section{A Kalman Filter Implementation of the Reduced-Rank MMSE Channel Estimator}
\label{sec:kalman}
As in the case of the Wiener filter, our goal is to reach a dimension reduction of the estimator. The conventional spatial dimension for a MIMO system with $N$ antennas is $N$, which can be large, especially for Massive MIMO Our goal is to reduce this dimensionality to $D$ where $D$ can be much smaller depending on the channel condition.
\subsection{Pre-beamforming Stage}
The pre-beamforming is applied in order to distinguish \textit{intra-group} signal of group $g$ users from other groups by suppressing the \textit{inter-group interference} while reducing the signaling dimension of $\mathbf{y}_n$ in (\ref{eqn:space_time_equivalent_model}). At the pre-beamforming stage, a $D$-dimensional vector $\mathbf{y}_n^{(g)}$ can be formed for all \textit{intra-cell} groups by a linear transformation through a matrix $\left(\mathbf{S}_D^{(g)}\right)^H$ as  
\begin{equation}
{\bf y}_n^{(g)} = \left({\bf\Psi}_n^{(g)}\right)^H {\bf h}_n^{(g)} + 
\left({\bf S}_D^{(g)}\right)^H \boldsymbol{\eta}_n^{(g)}
\label{eqn:SDHetan}
\end{equation}
where ${\bf\Psi}_n^{(g)} \triangleq {\bf x}_n^{(g)}\otimes {\bf S}_D^{(g)}$, and $\boldsymbol{\eta}_n^{(g)}$ is the \textit{inter-group} interference. Here, $\mathbf{S}_D^{(g)}$ can be regarded as $N \times D$ statistical pre-beamforming matrix that projects the $N$-dimensional received signal samples $\left\{\mathbf{y}_n\right\}$ in (\ref{eqn:multiuser_multipath_mimo_model}) 
on a suitable $D$-dimensional subspace in spatial domain.

\subsection{Reduced Rank Kalman Estimator} 
For channel variation in time, we adopt a state-space
model (\ref{eqn:SDHetan}) as a first-order stationary Gauss-Markov
process \cite{noh14}:
\begin{equation}
{\bf h}_n^{(g)} = \alpha {\bf h}_{n-1}^{(g)} +\sqrt{1-\alpha^2} 
{\bf b}_n^{(g)} .
\label{eqn:armodel}
\end{equation}
This is an autoregressive first order (AR(1)) model, with ${\bf b}_n$ being the input, which is also known as the disturbance. As we do not want the covariance of ${\bf h}_n^{(g)}$ change with $n$, we set 
\begin{equation}
\mathbb{E}\left\{{\bf h}_n^{(g)}\left({\bf h}_n^{(g)}\right)^H\right\}=
\mathbb{E}\left\{{\bf b}_n^{(g)}\left({\bf b}_n^{(g)}\right)^H\right\}
\triangleq {\bf R}_{\bf h}^{(g)}
\label{eqn:Eh_Eb_R_h_def}
\end{equation}
We focus on minimum mean square error
(MMSE) channel estimation based on the current and all previous
received training signals. The following MMSE estimate and its corresponding estimation error covariance are given by  
\begin{align}
\hat{\bf h}_{n|m}^{(g)} &\triangleq \mathbb{E}\left\{{\bf h}_{n}^{(g)}
\vert \mathbf{y}_k^{(g)},\;k=0,1,\ldots,m\right\} \nonumber \\
{\bf P}_{n|m}^{(g)} &\triangleq \mathbb{E}\left\{\left({\bf h}_{n}^{(g)}-
\hat{\bf h}_{n|m}^{(g)}\right)\left({\bf h}_{n}^{(g)}-\hat{\bf h}_{n|m}^{(g)}\right)^H \vert \mathbf{y}_k^{(g)},\;k=0,1,\ldots,m\right\}
\label{eqn:h_n_P_n_est}
\end{align}
Based on a standard derivation or a factor graph model, the MMSE estimates and the covariance matrices can be recursively computed based on Kalman filtering as shown in Table I. The Kalman filter description is standard. However, what we are implementing is a reduced-rank Kalman filter, that is to say, the rank of the Kalman gain matrix is significantly smaller than $N$ due to the statistical \textit{pre-beamformer}. To that end, the matrix ${\bf\Psi}_n^{(g)}$ is important. More specifically, we are seeking the subspace ${\bf S}_D^{(g)}$ that will achieve the rank reduction and the associated reduction in complexity.    

\begin{table}[htb]
\centering
\caption{Channel Estimation based on Reduced Rank Kalman Filtering after \textit{pre-beamformer} ${\bf S}_D^{(g)}$ for user group $g$, $g=1,\ldots,G$} 
\label{RR_Kalman}
\begin{tabular}{
  @{}
  >{\linespread{1.2}\selectfont}m{8.75cm}
  @{}
  m{8.75cm}
  @{}
}
\toprule 
\tableequation{
    \textrm{\underline{\textit{Initialization:}}} 
    \qquad \hat{\bf h}_{0| -1}^{(g)} = {\bf 0}
    \quad{\rm and}\quad {\bf P}_{0|-1}^{(g)}
     ={\bf R}_{\bf h}^{(g)} \label{h_init}
  } \\
\textbf{while } $n=0,1,\ldots$ \textbf{ do } \\
\tableequation{
    \quad {\bf\Psi}_n^{(g)} = {\bf x}_n^{(g)}\otimes {\bf S}_D^{(g)}
    \label{Phi_SD}
  } \\
\quad \underline{\textit{Measurement Update:}} \\
  \tableequation{
    \qquad \textrm{Innovation: }{\bf z}_n^{(g)} = 
    {\bf y}_n^{(g)} - \left({\bf\Psi}_n^{(g)}\right)^H 
    \hat{\bf h}_{n|n-1}^{(g)} \label{innovation}
  } \\
\qquad Innovation Covariance: \\
  \tableequation{
    \qquad \quad 
    {\bf E}_n^{(g)} =\left({\bf\Psi}_n^{(g)}\right)^H
    {\bf P}_{n|n-1}^{(g)}{\bf\Psi}_n^{(g)}+
    \left({\bf S}_D^{(g)}\right)^H{\bf R}_{\boldsymbol{\eta}}^{(g)}
    {\bf S}_D^{(g)} \label{innovation_cov}
  } \\
  \tableequation{
    \qquad \textrm{Kalman Gain: }
    {\bf K}_n^{(g)}={\bf P}_{n|n-1}^{(g)}{\bf\Psi}_n^{(g)}
    \left({\bf E}_n^{(g)}\right)^{-1} \label{Kalman_gain}
  } \\
  \tableequation{
    \qquad \textrm{A Posteriori State Estimate: } 
    \hat{\bf h}_{n|n}^{(g)}=\hat{\bf h}_{n|n-1}^{(g)}
    +{\bf K}_n^{(g)}{\bf z}_n^{(g)} \label{Apost_h_est}
  } \\
\qquad A Posteriori Estimate Covariance: \\ 
  \tableequation{
    \qquad \qquad 
    {\bf P}_{n|n}^{(g)}=\left[{\bf I}-{\bf K}_n^{(g)}
    \left({\bf\Psi}_{n}^{(g)}\right)^H\right]
    {\bf P}_{n|n-1}^{(g)} \label{Apost_cov}
  } \\
\quad \underline{\textit{Prediction:}} \\
  \tableequation{
    \qquad \textrm{A Priori State Estimate: }
    \hat{\bf h}_{n+1|n}^{(g)}=\alpha\hat{\bf h}_{n|n}^{(g)}    
    \label{Apriori_h_est}
  } \\
\qquad A Priori Estimate Covariance: \\
  \tableequation{
    \qquad \qquad 
    {\bf P}_{n+1|n}^{(g)} = \alpha^2 {\bf P}_{n|n}^{(g)} 
    + (1-\alpha^2) {\bf R}_{\bf h}^{(g)}
    \label{Apriori_cov}
  } \\
\textbf{end while} \\
\bottomrule
\end{tabular}
\end{table}
The new subspace ${\bf S}_D^H$ will be different than the Wiener filter formulation earlier \cite{guvensen16}. In particular, it will also be time-varying, or evolving in time. In what follows, we will develop this new filter. 
 
Note that the a posteriori and the a priori estimate covariances can be written as
\begin{align}
{\bf P}_{n|n}^{(g)} &={\bf P}_{n|n-1}^{(g)}-{\bf P}_{n|n-1}^{(g)}
{\bf\Psi}_n^{(g)} \left[\left({\bf\Psi}_n^{(g)}\right)^H 
{\bf P}_{n|n-1}^{(g)}{\bf\Psi}_n^{(g)} \right. \nonumber \\
&\qquad \left. +\left({\bf S}_D^{(g)}\right)^H
{\bf R}_{\boldsymbol{\eta}}^{(g)}{\bf S}_D^{(g)}\right]^{-1}
\left({\bf\Psi}_n^{(g)}\right)^H {\bf P}_{n|n-1}^{(g)} \label{Pnn} \\
{\bf P}_{n|n-1}^{(g)} &= \alpha^2 {\bf P}_{n-1|n-1}^{(g)} + 
\left(1-\alpha^2\right) {\bf R}_{\bf h}^{(g)}, 
\quad {\bf P}_{0|-1}^{(g)}={\bf R}_{\bf h}^{(g)}
\end{align}
where we can express ${\bf R}_{\bf h}^{(g)}$ in (\ref{eqn:Eh_Eb_R_h_def}) by using (\ref{eqn:mimo_multipath_correlations}) and (\ref{eqn:vector_def2}) in the following form  
\begin{equation}
{\bf R}_{\bf h}^{(g)} = \sum_{l=0}^{L_g-1} {\bf I}_{K_g}\otimes {\bf E}_{L_g,l}\otimes \rho_l^{(g)} {\bf R}_l^{(g)},
\label{eqn:R_h_g}
\end{equation}
where $\mathbf{E}_{L_g,l}$ is an $L_g \times L_g$ elementary diagonal matrix where all the entries except the $\left(l+1\right)^{th}$ diagonal one are zero.
By applying the matrix inversion lemma on (\ref{Pnn}), we get
\begin{equation}
\left({\bf P}_{n|n}^{(g)}\right)^{-1} = 
\left({\bf P}_{n|n-1}^{(g)}\right)^{-1} + 
{\bf\Psi}_n^{(g)} \left[\left({\bf S}_D^{(g)}\right)^H 
{\bf R}_{\boldsymbol{\eta}}{\bf S}_D^{(g)}\right]^{-1}
\left({\bf\Psi}_n^{(g)}\right)^H.
\label{eqn:Pnn-1}
\end{equation}

\subsection{Sequential Beam Design}
In this section, our goal is to find a good subspace (spanned by the columns of $\mathbf{S}_D^{(g)}$ matrix) on which the reduced dimensional instantaneous Kalman estimator can be realized as accurately as possible, so that a minimal performance compromise in the subsequent statistical signal processing operations after beamforming is provided. Different criteria can be used to design beam pattern as similar to one in \cite{guvensen16}. In this paper, we adopt error volume as the optimization criterion, namely, the minimization of $\det\left({\bf P}_{n|n}^{(g)}\right)$, and $\mathbf{S}_D^{(g)}$ is designed to minimize the error volume at each signaling interval for a given prediction error covariance ${\bf P}_{n|n-1}^{(g)}$ that is a function of all previous training signals.     

We consider to construct ${\bf S}_D^{(g)}$ optimally at beginning of each $M$ consecutive signaling interval where the symbol time $n=mM+u$ for $m=0,1,\ldots,$ and $0 \leq u \leq M-1$. Here, $M$ can be considered as the consecutive channel transmissions composed of a training period followed by a data transmission period as a block, and the \textit{pre-beamformer} is supposed to be updated at each of this interval.   
Assuming that the channel is almost stationary for $M$ consecutive symbol interval, one get the following equations in (\ref{eqn:Pnn-1-M-ver1}) and (\ref{eqn:Pnn-1-M-ver2}) together with (\ref{eqn:PnMn})
\begin{equation}
{\bf P}_{n+M|n}^{(g)} = \alpha^{2M} {\bf P}_{n|n}^{(g)} + 
\left(1 - \alpha^{2M}\right){\bf R}_{h}^{(g)}
\label{eqn:PnMn}
\end{equation}
where $\left(1 - \alpha^{2M}\right)$ should be approximately equal to 0 so that stationarity is ensured.
\begin{figure*}[htbp]
\begin{align}
\left({\bf P}_{n|n}^{(g)}\right)^{-1} &= 
\left({\bf P}_{n|n-M}^{(g)}\right)^{-1} + 
\sum_{m=0}^{M-1}
{\bf\Psi}_{n-m}^{(g)} \left[\left({\bf S}_D^{(g)}\right)^H 
{\bf R}_{\boldsymbol{\eta}}^{(g)}{\bf S}_D^{(g)}\right]^{-1}
\left({\bf\Psi}_{n-m}^{(g)}\right)^H \nonumber \\
&=\left({\bf P}_{n|n-M}^{(g)}\right)^{-1} +
\Big(\underbrace{\left[{\bf x}_n^{(g)} {\bf x}_{n-1}^{(g)}
\ldots {\bf x}_{n-M+1}^{(g)}\right]}_{\triangleq 
\displaystyle{\bf X}_n^{(g)}} \otimes {\bf S}_D^{(g)}\Big)
\left({\bf I}_M \otimes \left[\left({\bf S}_D^{(g)}\right)^H 
{\bf R}_{\boldsymbol{\eta}}^{(g)}{\bf S}_D^{(g)}\right]^{-1}\right)\left(\left[{\bf x}_n^{(g)} {\bf x}_{n-1}^{(g)}
\ldots {\bf x}_{n-M+1}^{(g)}\right] \otimes {\bf S}_D^{(g)}\right)^H 
\label{eqn:Pnn-1-M-ver1} \\
&=\left({\bf P}_{n|n-M}^{(g)}\right)^{-1}+
\left(\displaystyle{\bf X}_n^{(g)}
\left(\displaystyle{\bf X}_n^{(g)}\right)^H\right) \otimes
\left({\bf S}_D^{(g)}\left[\left({\bf S}_D^{(g)}\right)^H 
{\bf R}_{\boldsymbol{\eta}}^{(g)}{\bf S}_D^{(g)}\right]^{-1}
\left({\bf S}_D^{(g)}\right)^H\right)
\label{eqn:Pnn-1-M-ver2}
\end{align}
\end{figure*}

The eqn. (\ref{eqn:Pnn-1-M-ver1}) results in
\begin{align}
& \det\left({\bf P}_{n|n}^{(g)}\right) = \nonumber \\
& \frac{\det\left({\bf P}_{n|n-M}^{(g)}\right)}
{\det\left\{{\bf I} + \left({\bf I}_M \otimes \left[\left({\bf S}_D^{(g)}\right)^H {\bf R}_{\boldsymbol{\eta}}^{(g)}{\bf S}_D^{(g)}\right]^{-1}\right)
\left(\left(\displaystyle{\bf F}_n^{(g)}\right)^H 
{\bf P}_{n|n-M}^{(g)}\displaystyle{\bf F}_n^{(g)} 
\right)\right\}}.
\label{eqn:detPn}
\end{align}
where $\displaystyle{\bf F}_n^{(g)} \triangleq 
\displaystyle{\bf X}_n^{(g)} \otimes {\bf S}_D^{(g)}$.  
It is difficult to use (\ref{eqn:detPn}) to find the optimal ${\bf S}_D^{(g)}$ since it depends on the training data. We will use an approximation of (\ref{eqn:Pnn-1-M-ver2}) to get to a solution.
In an approximation similar to that performed to reach the stochastic gradient in the least mean square (LMS) algorithm, 
we will replace $\displaystyle{\bf X}_n^{(g)}
\left(\displaystyle{\bf X}_n^{(g)}\right)^H$ with its expected value $\mathbb{E}\left\{\displaystyle{\bf X}_n^{(g)}
\left(\displaystyle{\bf X}_n^{(g)}\right)^H\right\} = (E_s M) {\bf I}_{K_gL_g}$ to 
yield a block diagonal ${\bf P}_{n|n}^{(g)}$: 
\begin{align}
{\bf P}_{n|n}^{(g)} &= \left[\left({\bf P}_{n|n-M}
^{(g)}\right)^{-1} + (E_s M) \left({\bf I}_{K_gL_g} \otimes 
{\bf S}_D^{(g)}\right) \right. \nonumber \\
&\; \left. \left({\bf I}_{K_gL_g} \otimes \left[\left({\bf S}_D^{(g)}\right)^H {\bf R}_{\boldsymbol{\eta}}^{(g)}{\bf S}_D^{(g)}\right]^{-1}\right)\left({\bf I}_{K_gL_g} \otimes 
{\bf S}_D^{(g)}\right)^H\right]^{-1} \nonumber \\
&= {\bf P}_{n|n-M}^{(g)}-(E_s M){\bf P}_{n|n-M}^{(g)}
\left({\bf I}_{K_gL_g} \otimes 
{\bf S}_D^{(g)}\right) \nonumber \\
&\; \times \left[{\bf I}_{K_gL_g} \otimes 
\left({\bf S}_D^{(g)}\right)^H {\bf R}_{\boldsymbol{\eta}}^{(g)} {\bf S}_D^{(g)} + (E_sM)\left({\bf I}_{K_gL_g} 
\otimes {\bf S}_D^{(g)}\right)^H \right. \nonumber \\
&\; \left. {\bf P}_{n|n-M}^{(g)} \left({\bf I}_{K_gL_g} \otimes 
{\bf S}_D^{(g)}\right)\right]^{-1} 
\left({\bf I}_{K_gL_g} \otimes {\bf S}_D^{(g)}\right)^H
{\bf P}_{n|n-M}^{(g)}
\label{eqn:Pnn-1-M-approx}
\end{align}
Since initially ${\bf P}_{0|-1}^{(g)}={\bf R}_{\bf h}^{(g)}$ has block-diagonal structure given in (\ref{eqn:R_h_g}), ${\bf P}_{n|n}^{(g)}$ and 
${\bf P}_{n|n-M}^{(g)}$ can be expressed in the following form:  
\begin{equation}
{\bf P}_{n|m}^{(g)} = \sum_{l=0}^{L_g-1} {\bf I}_{K_g} \otimes {\bf E}_{L_g,l}\otimes {\bf A}_{n|m}^{l,(g)}.
\label{eqn:block_A_g_l}
\end{equation}

In order to construct the nearly optimal ${\bf S}_D^{(g)}$, first the following matrices in (\ref{eqn:block_snr_metric}) are constructed.   
\begin{figure*}[htbp]
\begin{align}
\left[{\bf I}_{K_gL_g} \otimes \left({\bf S}_D^{(g)}\right)^H 
{\bf R}_{\boldsymbol{\eta}}^{(g)}{\bf S}_D^{(g)}\right]^{-1}
\left[\left({\bf I}_{K_gL_g} \otimes {\bf S}_D^{(g)}
\right)^H {\bf P}_{n|m}^{(g)} \left({\bf I}_{K_gL_g} \otimes 
{\bf S}_D^{(g)}\right)\right] &= 
\sum_{l=0}^{L_g-1}{\bf I}_{K_g}\otimes {\bf E}_{L_g,l} \otimes 
{\bf SNR}_{n|m}^{l,(g)},\;
\textrm{ where we defined } \nonumber \\ 
{\bf SNR}_{n|m}^{l,(g)} &\triangleq \left[\left({\bf S}_D^{(g)}\right)^H {\bf R}_{\boldsymbol{\eta}}^{(g)}{\bf S}_D^{(g)}\right]^{-1}\left[\left({\bf S}_D^{(g)}\right)^H 
{\bf A}_{n|m}^{l,(g)}{\bf S}_D^{(g)}\right]
\label{eqn:block_snr_metric}
\end{align}
\end{figure*}
Then, by using (\ref{eqn:Pnn-1-M-approx}), (\ref{eqn:block_A_g_l}) and (\ref{eqn:block_snr_metric}), we get the following recursive relations in (\ref{eqn:SNRnn_l}), which starts with (\ref{eqn:SNRnn_init})
\begin{equation}
{\bf SNR}_{0|-m}^{l,(g)}= \left[\left({\bf S}_D^{(g)}\right)^H {\bf R}_{\boldsymbol{\eta}}^{(g)}{\bf S}_D^{(g)}\right]^{-1}
\left[\rho_l^{(g)}\left({\bf S}_D^{(g)}\right)^H 
{\bf R}_l^{(g)}{\bf S}_D^{(g)}\right]
\label{eqn:SNRnn_init}
\end{equation}
for $m=1,\ldots,M$. 
\begin{figure*}[htbp]
\begin{align}
{\bf SNR}_{n|n}^{l,(g)} &= {\bf SNR}_{n|n-M}^{l,(g)} - (E_sM) 
{\bf SNR}_{n|n-M}^{l,(g)} \left[{\bf I}_D + (E_sM)
{\bf SNR}_{n|n-M}^{l,(g)}\right]^{-1}{\bf SNR}_{n|n-M}^{l,(g)}
\nonumber \\
{\bf SNR}_{n+M|n}^{l,(g)} &= \alpha^{2M} {\bf SNR}_{n|n}^{l,(g)} + 
\left(1 - \alpha^{2M}\right){\bf SNR}_{0|-M}^{l,(g)},\;
l=0,\ldots,L_g-1.
\label{eqn:SNRnn_l}
\end{align}
\end{figure*}
Finally, the following error volume expression is obtained after using (\ref{eqn:Pnn-1-M-approx}) and the matrix inversion lemma: 
\begin{align}
\det\left({\bf P}_{n|n}^{(g)}\right)&= 
\frac{\det\left({\bf P}_{n|n-M}^{(g)}\right)}
{\det\left({\bf I}+(E_sM)\sum_{l=0}^{L_g-1}{\bf I}_{K_g} 
\otimes {\bf E}_{L_g,l} \otimes {\bf SNR}_{n|n-M}^{l,(g)}\right)}.
\label{eqn:detPnn_approx}
\end{align}

\underline{\textit{Proposition:}} Given all previous training signals, the pre-beamformer ${\bf S}_D^{(g)}$ at $n^{th}$ signaling interval minimizing 
$\det\left({\bf P}_{n|n}^{(g)}\right)$ in (\ref{eqn:detPnn_approx}) is given by a proper scaled version of the $D$ dominant \textit{generalized eigenvectors} of ${\bf R}_l^{(g)}$ and ${\bf R}_{\boldsymbol{\eta}}^{(g)}$
\cite{guvensen16}. Under this pre-beamformer design, all three matrices ${\bf SNR}_{n|n}^{l,(g)}$, ${\bf SNR}_{n|n-M}^{l,(g)}$ and ${\bf SNR}_{0|-m}^{l,(g)}$ are simultaneously diagonalizable, i.e., given these \textit{generalized eigenvectors} that diagonalize ${\bf SNR}_{0|-m}^{l,(g)}$ initially, the ${\bf SNR}_{n|n}^{l,(g)}$ and ${\bf SNR}_{n|n-M}^{l,(g)}$ are also diagonalizable for every $n$ from the recursive relation in (\ref{eqn:SNRnn_l}). This sequential beamspace construction algorithm is summarized in Algorithm 1.     

\begin{algorithm}
\caption{Sequential Pre-beamformer Construction}
\begin{algorithmic} 
\REQUIRE Obtain the \textit{generalized eigenvectors} of ${\bf R}_l^{(g)}$ and ${\bf R}_{\boldsymbol{\eta}}^{(g)}$ by solving 
$\rho_l^{(g)}{\bf R}_l^{(g)}{\bf V}^l={\bf R}_{\boldsymbol{\eta}}^{(g)}
{\bf V}^l\boldsymbol{\Lambda}^l,\;l=0,\ldots,L_g-1$, \\
$\boldsymbol{\Lambda}^l_0 \leftarrow \boldsymbol{\Lambda}^l$, \\
\textbf{while } $m=0,1,2,\ldots$ \textbf{ do } \\
The pre-beamformer is in the form \cite{guvensen16}: 
\begin{equation}
\mathbf{S}_D^{(g)} \triangleq
\left[\begin{array}{cccc}
\mathbf{S}_D^{(g)}(0) & \mathbf{S}_D^{(g)}(1) & \cdots & \mathbf{S}_D^{(g)}(L_g-1)
\end{array}\right]_{N \times D} 
\label{eqn:S_D_l_approx}
\end{equation}
${\bf SNR}_{0|-M}^{l,(g)}$ is diagonalized with $N \times d_l$ $\mathbf{S}_D^{(g)}(l)$ constructed by the columns of ${\bf V}^l$ corresponding to the $d_l$ dominant \textit{generalized eigenvalues} for 
$\sum_{l=0}^{L_g-1}d_l=D$.\\
In order to find the set of column indices of ${\bf V}^l$ showing the $d_l$ largest eigenvalues at each signaling interval $m$, namely, $\left\{\mathcal{I}_m^l\right\}$, solve the following  
\begin{equation}
\left\{\mathcal{I}_m^l\right\}=
\underset{\left\{\mathcal{I'}_m^l\right\}_{l=0}^{L_g-1}}{\operatorname{argmax}}\;
\underbrace{\det\left({\bf I}+(E_sM)\sum_{l=0}^{L_g-1}{\bf I}_{K_g} \otimes {\bf E}_{L_g,l} \otimes \boldsymbol{\Lambda}^l
_{\left\{\mathcal{I'}_m^l\right\}}\right)}_{\mathfrak{F}
\left(\left\{\mathcal{I'}_m^l\right\}_{l=0}^{L_g-1}\right)}
\label{eqn:det_metric_opt_GE_set}
\end{equation}
where $|\left\{\mathcal{I}_m^l\right\}|=d_l$, and 
\begin{equation}
\operatorname{max}\;
\mathfrak{F}\left(\left\{\mathcal{I}_m^l\right\}_{l=0}^{L_g-1}\right)=\prod_{l=0}^{L_g-1}
\prod_{i=1}^{d_l}\left(1+(E_s M) \lambda^l_i\right)^{K_g}
\label{eqn:det_metric_opt_GE_value}
\end{equation}
by assuming that ${\bf SNR}_{n|n-M}^{l,(g)}$ have nearly orthogonal eigenspaces for different $l$, and $\lambda^l_i$'s are the diagonal entries of $\boldsymbol{\Lambda}^l$ whose indices are given by the set 
$\left\{\mathcal{I}_m^l\right\}$. \\
\qquad \emph{Update:}  
\begin{equation}
\mathbf{S}_D^{(g)}(l) \leftarrow 
{\bf V}^l_{\left\{\mathcal{I}_m^l\right\}},\;
l=0,\ldots,L_g-1
\label{eqn:S_D_update}
\end{equation}
\begin{align}
\boldsymbol{\Lambda}^l_{\left\{\mathcal{I}_m^l\right\}} &\leftarrow
\boldsymbol{\Lambda}^l_{\left\{\mathcal{I}_m^l\right\}} - (E_sM)
\left(\boldsymbol{\Lambda}^l_{\left\{\mathcal{I}_m^l\right\}}\right)^2.\bigg/\left({\bf I}_D + (E_sM)
\boldsymbol{\Lambda}^l_{\left\{\mathcal{I}_m^l\right\}}\right)
\nonumber \\[10pt]
\boldsymbol{\Lambda}^l &\leftarrow \alpha^{2M} \boldsymbol{\Lambda}^l + \left(1 - \alpha^{2M}\right)
\boldsymbol{\Lambda}^l_0.
\label{eqn:lambda_l_update}
\end{align}
\textbf{end while} 
Here, $.\bigg/$ denotes the elementwise division and $\boldsymbol{\Lambda}^l_{\left\{\mathcal{I}_m^l\right\}}$ shows the diagonal elements of
$\boldsymbol{\Lambda}^l$ whose indices given by the set $\left\{\mathcal{I}_m^l\right\}$ at the $m^{th}$ signaling interval.   
\end{algorithmic}
\end{algorithm}

\section{Numerical Results and Discussion}
\label{sec:Sim_Results}
In this section, we provide some numerical results to evaluate the performance of the reduced rank Kalman Estimator for the proposed pattern design of the \textit{pre-beamformer}. Throughout the demonstrations, we consider a massive MIMO system with uplink training in TDD mode where a BS is equipped with a uniform linear array (ULA) of $N=100$ antenna elements along the y-axis% 
\footnote{Although the proposed estimators are valid for an arbitrary array structure in this paper, ULA is considered for the ease of exposition only.}, and each of $K$ users has a single receive antenna. 

In the studied scenario, $K$ users were clustered into eight groups ($G=8$), and each UT is assumed to be located at a specific azimuth angle $\theta$ along the ring centered at the origin in the x-y plane. The channel covariance matrix of each group is specified with the center azimuth angle $\theta$ (AoA), and can be calculated in a similar way to the ones in \cite{adhikary13,kim15}. In the simulations, our focus is on the channel estimation accuracy of the intended group 
$g$ with $3$ MPCs, i.e., $L_g=3$. The first two MPCs of group $g$ stem from a azimuth angular sector $[-1^\circ,1^\circ]$ for delays at $l=0,1$, and the angular sector of the last MPC at $l=2$ of $g$ is given as $[5^\circ,7^\circ]$ in azimuth. We assume two users served simultaneously for group $g$, i.e., $K_g=2$. Each of the other $7$ groups (interfering with the intended one) consists of three users, i.e., $K_{g'}=3,\;g' \neq g$ and these users have $3$ MPCs whose angular sectors have same supports of AoA ($L_{g'}=3,\;g' \neq g$) given by 
$[-29 , -26]$, $[-21 , -19]$, $[-12 , -9]$, $[-5.5 , -3.5]$, 
$[9.5 , 12.5]$, $[15 , 17]$, $[24 , 27]$ in azimuth respectively. The noise power is set as $N_0=1$ so that all dB power values are relative to $1$. %In TDD mode, \textit{inter-group} users do not need to be synchronized, and even are allowed to use the same sequences during uplink training mode. 
\textit{Intra-group} users (of the intended group) use non-orthogonal training waveforms composed of $T$ chips, and these are obtained by truncating length-63 \textit{Kasami} codes by simply choosing the first $T$ chips (training length) of last $K_g$ \textit{Kasami} sequences without any optimization. 
%\footnote{There are more efficient approaches (other than the truncation of Kasami codes) yielding waveforms with better cross- and auto-correlation properties and minimizing (\ref{eqn:nmse_S_D_opt}), but training optimization is beyond the scope of this study}
%Then, the training matrix in (\ref{eqn:complete_training_matrix}) can be constructed to be exploited by the BS during CSI acquisition period. 

The trace of the estimation error covariance matrix given by (\ref{Apost_cov}) for the extended channel vector of group $g$ users in (\ref{eqn:vector_def2}) to compare the performance of RR Kalman estimators based on different pre-beamformers. 
The covariance matrix of the \textit{inter-group interference} is evaluated by (\ref{eqn:InterGroupInterference}) when the angular sector of each group is provided.
  
In this paper, we compare the performance of dimension reduction based on Generalized Eigendecomposition in Algorithm 1 (shown to be nearly optimal under some realistic assumptions) with that of the conventional subspace composed of the first $D$ dominant eigenvectors of $\sum_{l=0}^{L_g-1}\rho_l^{(g)}\mathbf{R}_l^{(g)}$ in (\ref{eqn:mimo_multipath_correlations}). We call this conventional beamspace as discrete Fourier transform (DFT) beamspace because the eigenvectors of the spatial correlation matrix of the ULA channel are well-approximated by the columns of the $N \times N$ unitary DFT matrix whose indices correspond to the support of the Fourier transform of the spatial correlation function (owing to the Szeg\"o's asymptotic theory) \cite{nam14} depending on the angular sector of group $g$ UTs. This conventional beamspace is known to be information preserving for spatially white interference case, and thus, widely used in practical \textit{hybrid beamforming} applications, where the beamforming in RF analog domain can be implemented by simple phase shifters \cite{liu14}.  

\begin{figure}[htbp]
\centering
\epsfig{file=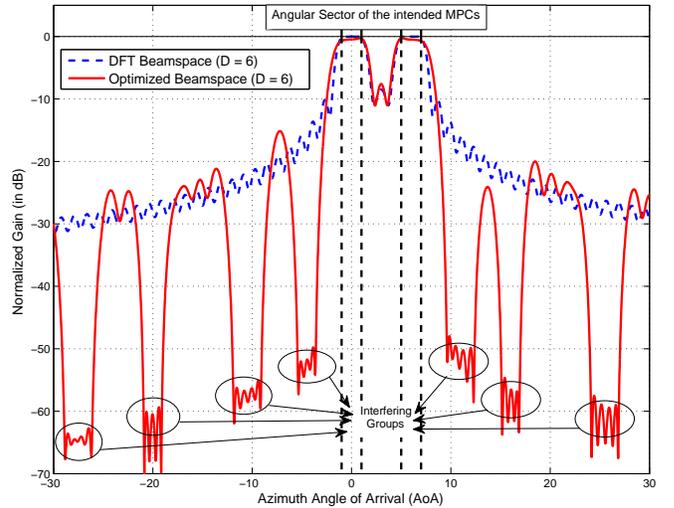,width=0.5\textwidth}
\caption{Beam pattern of different pre-beamformers}  
\label{fig:Pattern_Dim6_equalINR_SNR40}
\vskip -1em
\end{figure}

In Figure \ref{fig:Pattern_Dim6_equalINR_SNR40}, the beam patterns created by the generalized eigenvector beamspace (GEB) and DFT beamspaces are depicted for $D=6$ at $snr=30$ (dB). The GEB is designed based on the AoA support of the intended group $g$ for $l=0,1,2$ while taking the angular locations of the interfering groups into account. The \textit{inter-group} users signal are assumed to have same power level with that of the intended group. As can be seen from the figure, the GEB tries to create deep nulls at the angular locations of interfering UTs, whereas the conventional pre-beamformer only tries to maximize the captured power of the intended group MPCs for a given dimension. It is expected that as the number of BS antennas increases, the eigenspaces of each group are approximately orthogonal. However, the number of transmit antennas is finite in practice, and there always exists some overlap among the virtual angular sectors of each group which leads to a leakage to the intended group signal. Therefore, as it will be shown later, the accuracy of the channel estimation realized on the reduced dimensional subspace, spanned by the conventional DFT beamspace, is considerably lost due to the residual \textit{inter-group interference} after pre-beamforming. On the other hand, the GEB suppresses the \textit{inter-group interference} while allowing the MPCs of the intended group to pass with a negligible distortion. 
%so that the subsequent processing in reduced dimensions, here the instantaneous CSI estimation, can be carried out as accurately as possible.     

\begin{figure}[htbp]
\centering
\epsfig{file=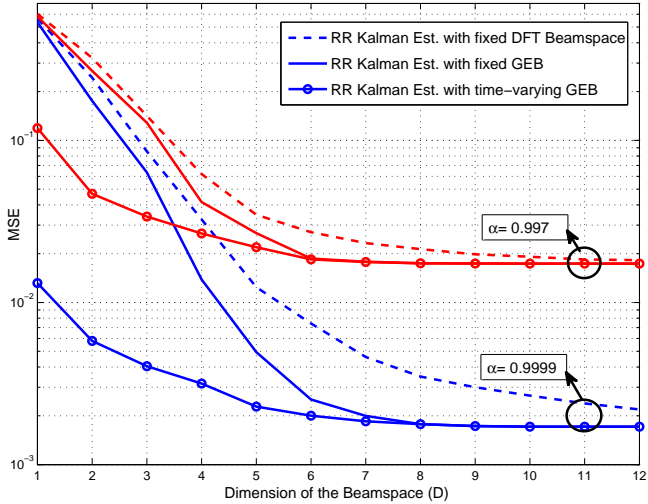,width=0.475\textwidth}
\caption{MSE vs dimension for time-varying channels after $50$ training symbol transmission at $snr= 30$ dB, $M=5$}  
\label{fig:mean_error_kalman_mse_vs_dim}
\vskip -1em
\end{figure}

In Figure \ref{fig:mean_error_kalman_mse_vs_dim}, the average mean square error (MSE) values given by $\operatorname{Tr}{\left\{\mathbf{P}_{n|n}^{(g)}\right\}}/K_g$ as a function of the dimension of the spatial domain pre-beamformer ($D$) are depicted for both fixed and time-varying pre-beamformer cases. In fixed GEB case, the pre-beamformer is optimized based on the initial uncertainties, i.e., initial channel covariance, and kept fixed during whole transmission period while realizing RR Kalman estimator. Whereas, in time-varying beamspace construction, the proposed sequential design algorithm (Algorithm 1) is used. This algorithm, taking the varying Kalman prediction errors in each eigen directions into account so as to reduce error volume, results in much better dimension reduction capability, i.e., for a given dimension, time-varying beamspace construction yields much lower channel estimation errors compared to fixed design alternatives.    

\begin{figure}[htbp]
\centering
\epsfig{file=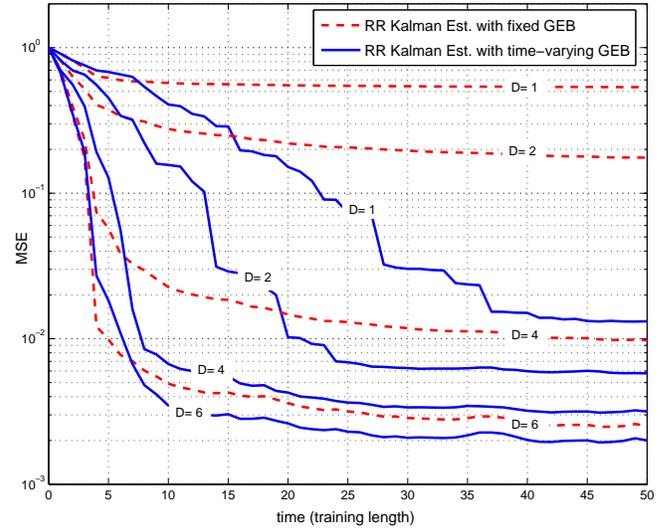,width=0.475\textwidth}
\caption{MSE vs training length for time-varying channel with 
$\alpha=0.9999$ at $snr= 30$ dB, $M=5$}  
\label{fig:mean_error_kalman_mse_vs_time}
\vskip -1em
\end{figure}

In Figure \ref{fig:mean_error_kalman_mse_vs_time}, the average mean square error (MSE) values as a function of training length are demonstrated for different beamspace dimensions. Fixed statistical pre-beamformer case show much inferior performance to time-varying case even with the use of GEB. Even though the convergence is slower for lower dimensional pre-beamformers in sequential pre-beamformer design, it achieves a much lower steady state estimation error.   

\section{Conclusions}
\label{sec:Conclusion}
In two-stage beamforming, a new algorithm for statistical pre-beamformer design together with Kalman filtering based channel estimator is proposed for massive MIMO transmission employing SC in frequency-selective fading. The proposed algorithm yields a nearly optimal pre-beamformer whose beam pattern is designed sequentially with low complexity by taking the user-grouping into account, and exploiting the properties of Kalman filtering and associated prediction error covariance matrices. The resulting design, based on the second order statistical properties of the channel, generates much lower mean square error values compared to conventional beamforming techniques.   

\section*{Acknowledgement}
This work was supported in part by the National Science
Foundation under grant no. 1547155, and 
ASELSAN Corp., Ankara, Turkey. 

\bibliographystyle{ieeetr}
\bibliography{SC_MIMO_MMW}

\begin{thebibliography}{10}

\bibitem{andrews14}
J.~G. Andrews, S.~Buzzi, W.~Choi, S.~V. Hanly, A.~Lozano, A.~C.~K. Soong, and
  J.~C. Zhang, ``What will 5{G} be?,'' {\em {IEEE} J. Sel. Areas Commun.},
  vol.~32, pp.~1065--1082, Jun. 2014.

\bibitem{swindlehurst14}
L.~Lu, G.~Y. Li, A.~L. Swindlehurst, A.~Ashikhmin, and R.~Zhang, ``An overview
  of massive {MIMO}: Benefits and challenges,'' {\em {IEEE} J. Sel. Areas
  Commun.}, vol.~8, pp.~742--758, Oct. 2014.

\bibitem{ashikhmin11}
J.~Jose, A.~Ashikhmin, T.~L. Marzetta, and S.~Vishwanath, ``Pilot contamination
  and precoding in multi-cell {TDD} systems,'' {\em {IEEE} Trans. Wireless
  Commun.}, vol.~10, pp.~2640--2651, Aug. 2011.

\bibitem{adhikary13}
A.~Adhikary, J.~Nam, J.~Y. Ahn, and G.~Caire, ``Joint spatial division and
  multiplexing: The large-scale array regime,'' {\em {IEEE} Trans. Inf.
  Theory}, vol.~59, pp.~6441--6463, Oct. 2013.

\bibitem{nam14}
J.~Nam, A.~Adhikary, J.~Y. Ahn, and G.~Caire, ``Joint spatial division and
  multiplexing: Opportunistic beamforming, user grouping and simplified
  downlink scheduling,'' {\em IEEE J. Sel. Topics Signal Process.}, vol.~8,
  pp.~876--890, Oct. 2014.

\bibitem{kim15}
D.~Kim, G.~Lee, and Y.~Sung, ``Two-stage beamformer design for massive {MIMO}
  downlink by trace quotient formulation,'' {\em {IEEE} Trans. Commun.},
  vol.~63, pp.~2200--2211, Jun. 2015.

\bibitem{adhikary14}
A.~Adhikary, E.~A. Safadi, M.~K. Samimi, R.~Wang, G.~Caire, T.~S. Rappaport,
  and A.~F. Molisch, ``Joint spatial division and multiplexing for mm-wave
  channels,'' {\em {IEEE} J. Sel. Areas Commun.}, vol.~32, pp.~1239--1255, Jun.
  2014.

\bibitem{swindlehurst16}
L.~You, X.~Gao, A.~L. Swindlehurst, and W.~Zhong, ``Channel acquisition for
  massive {MIMO}-{OFDM} with adjustable phase shift pilots,'' {\em {IEEE}
  Trans. Signal Process.}, vol.~64, pp.~1461--1476, Mar. 2016.

\bibitem{you15}
L.~You, X.~Gao, X.~G. Xia, N.~Ma, and Y.~Peng, ``Pilot reuse for massive {MIMO}
  transmission over spatially correlated rayleigh fading channels,'' {\em
  {IEEE} Trans. Wireless Commun.}, vol.~14, pp.~3352--3366, Jun. 2015.

\bibitem{rappaport13}
T.~S. Rappaport, S.~Sun, R.~Mayzus, H.~Zhao, Y.~Azar, K.~Wang, G.~N. Wong,
  J.~K. Schulz, M.~Samimi, and F.~Gutierrez, ``Millimeter wave mobile
  communications for 5{G} cellular: It will work!,'' {\em IEEE Access}, vol.~1,
  pp.~335--349, May 2013.

\bibitem{utschick05}
F.~A. Dietrich and W.~Utschick, ``Pilot-assisted channel estimation based on
  second-order statistics,'' {\em {IEEE} Trans. Signal Process.}, vol.~53,
  pp.~1178--1193, Mar. 2005.

\bibitem{noh14}
S.~Noh, M.~D. Zoltowski, Y.~Sung, and D.~J. Love, ``Pilot beam pattern design
  for channel estimation in massive mimo systems,'' {\em IEEE J. Sel. Topics
  Signal Process.}, vol.~8, no.~5, pp.~787--801, 2014.

\bibitem{guvensen16}
G.~M. Guvensen and E.~Ayanoglu, ``A generalized framework on beamformer design
  and {CSI} acquisition for single-carrier massive {MIMO} systems in millimeter
  wave channels,'' {\em arXiv preprint arXiv:1607.01436}, 2016.

\bibitem{liu14}
A.~Liu and V.~Lau, ``Phase only {RF} precoding for massive {MIMO} systems with
  limited {RF} chains,'' {\em {IEEE} Trans. Signal Process.}, vol.~62,
  pp.~4505--4515, Sep. 2014.

\end{thebibliography}
\end{document}